\documentstyle[preprint,aps,epsf,tighten]{revtex}	
\begin{document}
%
%
\preprint{$
\begin{array}{l}
\mbox{BA-03-21}\\
\mbox{FERMILAB-Pub-03/405-T}\\
\mbox{hep-ph/0312224}\\
\mbox{December 2003}\\[0.3in]
\end{array}
$}
\title{Leptogenesis in the type III seesaw mechanism}
\author{Carl H. Albright}
\address{Department of Physics, Northern Illinois University, DeKalb, IL
60115\\
       and\\
Fermi National Accelerator Laboratory, P.O. Box 500, Batavia, IL
60510\footnote{electronic address: albright@fnal.gov}}
\author{S.M. Barr}
\address{Bartol Research Institute,
University of Delaware, Newark, DE 19716\footnote{electronic address:
smbarr@bartol.udel.edu}}
\maketitle
\thispagestyle{empty}
\begin{abstract}
It is shown that the type III seesaw mechanism proposed recently can have
certain advantages over the conventional (or type I) seesaw mechanism for
leptogenesis. In particular a resonant enhancement of leptogenesis via 
heavy quasi-Dirac right-handed neutrino pairs can occur without a special 
flavor form or ``texture" of the mass matrices being assumed. Some of the 
requirements for neutrino mixing and leptogenesis are effectively decoupled.
\end{abstract}
%
\pacs{PACS numbers: 12.15Ff, 12.10.Dm, 12.60.Jv, 14.60.Pq}
%
%

\section{INTRODUCTION}
Recently there has been much interest in the idea of leptogenesis through the 
decay of heavy right-handed neutrinos 
\cite{leptogen1,leptogen2,leptogen3,leptogen4}. In part, 
this interest is due to the fact that such a scenario links the origin 
of matter to the masses and mixings of neutrinos, about which we are now 
learning from experiment. Most analyses of leptogenesis have 
focussed on the conventional seesaw mechanism for neutrino mass 
\cite{seesaw}, which is
sometimes called the ``type I seesaw mechanism" to distinguish it from 
the ``type II seesaw mechanism" discussed in \cite{type2} involving
Higgs triplets. An analysis of thermal 
leptogenesis in models where the type II mechanism operates has been given in 
Ref. \cite{leptotype2}. 

In a recent paper \cite{type3} it was shown that in a wide class of
models based on $SO(10)$ (or larger groups) a somewhat different seesaw
mechanism can operate that has several interesting features. In particular,
this ``type III seesaw" may have less difficulty in reproducing realistic
neutrino masses and mixings than does the conventional type I seesaw. At
least, this was found to be the case in a particular, rather typical,
realistic $SO(10)$ model of quark and lepton masses \cite{ab}. In the 
present paper we show 
that the type III seesaw mechanism also has some significant advantages 
for leptogenesis.

In realistic SUSY GUT models, it is typically found that for there to be
sufficient leptogenesis there must be resonant enhancement \cite{leptogen3}
coming from two superheavy ``right-handed" neutrinos 
(call them $N_1$ and $N_2$) 
forming what we shall call a ``nearly degenerate pseudo-Dirac pair". 
What we mean by this is that they have a mass matrix of the form
\begin{equation}
(N_1, N_2) \left( \begin{array}{cc} \Delta_{11} & M \\ M & \Delta_{22}
\end{array} \right) \left( \begin{array}{c} N_1 \\ N_2 \end{array} \right),
\end{equation}

\noindent
with $\Delta_{ii} \ll M$. This is equivalent, upon diagonalization, to two
Majorana neutrinos $N_+ \cong (N_1 + N_2)/\sqrt{2}$ and $N_- \cong
(N_1 - N_2)/\sqrt{2}$, with masses approximately equal to
$M + \frac{1}{2}(\Delta_{11} + \Delta_{22})$
and $-M + \frac{1}{2}(\Delta_{11} + \Delta_{22})$ respectively. Then the
lepton asymmetry produced by the decays of these heavy neutrinos will
be enhanced by a factor of $I \cong \frac{M}{2(\Delta_{11} + \Delta_{22})}$. 
In many models this factor must be very large to obtain sufficient 
leptogenesis \cite{resneeded}.

In the conventional or type I seesaw, the requirement for resonant enhancement
that there be such a nearly degenerate right-handed neutrino pair places a 
very strong constraint on the form of $M_R$, the mass matrix of the 
right-handed neutrinos. The matrix $M_R$ also appears in
the type I seesaw formula for the mass matrix of the
observed light neutrinos: $M_{\nu} = - M_N M_R^{-1} M_N^T$. (Here and 
henceforth $M_N$ denotes
the Dirac mass matrix that links the left-handed and right-handed neutrinos.)
The leptogenesis constraints on $M_R$ can therefore clash with what is
needed to give a realistic pattern of light neutrino masses and mixings.
Indeed, in Ref. \cite{abtype1} it was found that this does happen in a 
typical realistic $SO(10)$ model of quark and lepton masses. 

In the type III seesaw, the right-handed neutrino mass matrix does not 
enter in the same way into the formula for $M_{\nu}$ as in the type I seesaw. 
Consequently, the issues of leptogenesis and of realistic light neutrino 
masses and mixings are not so tightly coupled, and the clash 
referred to above can easily be avoided. Moreover, nearly degenerate
pseudo-Dirac pairs of heavy neutrinos arise very simply and naturally
in the type III seesaw mechanism, as will be seen. Their existence does 
not impose any
special form in flavor space on the mass matrices involved. In fact, in the
type III seesaw mechanism there are not (as in type I) three
superheavy Majorana neutrinos, but rather six superheavy Majorana neutrinos, 
which form three pseudo-Dirac pairs. One or more of these pairs can be highly 
degenerate without any special ``textures" or forms.

\section{Type III seesaw mechanism}

In this section we shall briefly review the type III seesaw mechanism. More 
details can be found in Ref. \cite{type3}.
The type III seesaw mechanism arises in $SO(10)$ models in which the
breaking of $B-L$ is done by spinor ${\bf 16}_H + \overline{{\bf 16}}_H$
Higgs fields rather than by rank-five tensor ${\bf 126}_H + 
\overline{{\bf 126}}_H$ Higgs fields. (The generalization to larger groups, 
like $E_6$ is straightforward.) In such a case, the right handed neutrino
masses must come from an effective $d=5$ operator of the form
\begin{equation}
O_{eff} = \lambda_{ijab} ({\bf 16}_i {\bf 16}_j) \overline{{\bf 16}}^a_H 
\overline{{\bf 16}}^b_H/M,
\end{equation}

\noindent
The ${\bf 16}_i$, $i=1,2,3$ are the three families of light quarks 
and leptons. The indices $a$ and $b$ on the Higgs multiplets indicate that
there can be more than one $\overline{{\bf 16}}_H$. We will denote by
${\bf p}({\bf q})$ an $SU(5)$ ${\bf p}$ representation contained in an
$SO(10)$ ${\bf q}$ representation. Thus, the left-handed neutrinos $\nu_i$ 
are contained in $\overline{{\bf 5}}({\bf 16}_i)$, while the left-handed 
antineutrinos $N^c_i$ (the conjugates of the right-handed neutrinos) are 
contained in ${\bf 1}({\bf 16}_i)$.
The $N^c_i$ obtain superlarge mass from $O_{eff}$
when the ${\bf 1}(\overline{{\bf 16}}^a_H)$ obtain GUT-scale vacuum expectation
values (VEVs): $\langle {\bf 1}(\overline{{\bf 16}}^a_H) \rangle \equiv 
\Omega_a$. Thus $O_{eff}$ gives a term of the form $N^c_i N^c_j (\lambda_{ijab}
\Omega_a \Omega_b/M)$.

The operator $O_{eff}$
must arise from integrating out some superheavy states, as shown in Fig. 1. 
For simplicity, we assume that the states integrated out are $SO(10)$ 
singlets, though what representations of
$SO(10)$ they are in does not really matter for the later discussion.
Suppose that there are three such singlets, one for each family,
denoted $S_i \equiv {\bf 1}_i$.
The terms needed in the Yukawa superpotential to give $O_{eff}$ from 
the diagram of Fig. 1 are
\begin{equation}
W_{RH\nu} = F^a_{ij} ({\bf 16}_i {\bf 1}_j) \overline{{\bf 16}}^a_H +
{\cal M}_{ij} {\bf 1}_i {\bf 1}_j.
\end{equation}

\noindent
There must also be a Dirac neutrino mass term 
$(M_N)_{ij} \nu_i N^c_j$. In a realistic model $M_N$
would typically come from a combination of several operators at the $SO(10)$
level. For our purposes, the origin of $M_N$ does not matter.

The crucial point for the type III seesaw mechanism is that the weak-doublet 
Higgs fields contained in the $\overline{{\bf 16}}^a_H$ can get weak-scale
VEVs: $|\langle {\bf 5}(\overline{{\bf 16}}^a_H \rangle| \equiv u_a$.
These VEVs directly couple the left-handed neutrinos $\nu_i$
to the singlets $S_i$.
The full mass matrix of neutral leptons has the form:
\begin{equation}
W_{neut} = (\nu_i, N^c_i, S_i) \left( \begin{array}{ccc}
0 & (M_N)_{ij} & F^a_{ij} u_a \\ (M_N)_{ji} & 0 & F^a_{ij} \Omega_a \\
F^a_{ji} u_a & F^a_{ji} \Omega_a & {\cal M}_{ij} \end{array} \right) 
\left( \begin{array}{c} \nu_j \\ N^c_j \\ S_j \end{array} \right).
\end{equation}

\noindent
The matrix ${\cal M}_{ij}$ and the VEVs $\Omega_a$ are superheavy, while the
mass matrix $M_N$ and the VEVs $u_a$ are of order the weak scale. 
In the special case where there is only one $\overline{{\bf 16}}_H$
the situation becomes very simple, for then the $\nu S$ block ($= F_{ij} u$) is
proportional to the $N^c S$ block ($= F_{ij} \Omega$). Consequently a rotation 
of $\nu_i$ and $N^c_i$ by an angle $\tan^{-1}(u/\Omega)$ makes the 
$\nu S$ and $S \nu$ blocks exactly vanish and induces in the $\nu \nu$ block 
the entry 
$-(M_N + M_N^T) u/\Omega$. The rest of the matrix has the standard ``double
seesaw" form. The total effective mass matrix for the three light neutrinos 
(neglecting terms subleading in $M_W/M_{GUT}$) is
\begin{equation}
M_{\nu} = - M_N M_R^{-1} M_N^T - (M_N + M_N^T) \frac{u}{\Omega},
\end{equation}

\noindent
where 
\begin{equation}
M_R = (F \Omega) {\cal M}^{-1} (F^T \Omega).
\end{equation}

In the general case where there is more than one $\overline{{\bf 16}}_H$,
the light neutrino mass matrix has the form
\begin{equation}
M_{\nu} = - M_N M_R^{-1} M_N^T - (M_N H + H^T M_N^T) \frac{u}{\Omega},
\;\; H \equiv (F' F^{-1})^T,
\end{equation}

\noindent
where $\Omega \equiv (\sum_a \Omega_a^2)^{1/2}$, $u \equiv 
(\sum_a u_a^2)^{1/2}$,
$F_{ij} \equiv \sum_a F^a_{ij} 
\Omega_a/\Omega$, and $F'_{ij} \equiv \sum_a F^a_{ij} u_a/u$. 

The first term in Eq. (7) is the usual type I seesaw contribution. The 
second term is the type III seesaw contribution. If the elements of the
matrix ${\cal M}_{ij}$ are small compared to those of 
$F^a_{ij} \Omega_a$, then it is easy to see from Eqs. (6) and (7)
that the type I contribution
becomes negligible compared to the type III contribution. We shall be
interested in this case, which is also favorable for leptogenesis, as will 
be seen.

In the limit that ${\cal M}_{ij} = 0$ one 
sees from Eq. (4) that the superheavy neutrinos have simply 
the mass term $F_{ij} \Omega (N^c_i S_j)$. That is, the $N^c_i$ and $S_i$ pair
up to form three Dirac neutrinos. On the other hand, if
${\cal M}_{ij}$ is small (compared to $F_{ij} \Omega$) but not zero, then 
these three Dirac neutrinos get slightly split into six eigenstates 
forming three nearly degenerate pseudo-Dirac
neutrinos. It is this fact that can be exploited to enhance leptogenesis.

\section{Leptogenesis in a realistic type III model}

Rather than attempting to study leptogenesis in type III seesaw models in 
complete generality, it may be more illuminating to study it in a specific 
realistic model of quark and lepton (including neutrino) masses and mixings.
We shall use the $SO(10)$ model of Ref. \cite{ab}.

One of the attractive features of $SO(10)$ unification (or unification based
on larger groups) is that the unified symmetry relates the Dirac mass matrix 
of the neutrinos, $M_N$, to the mass matrices of the up quarks, down quarks and
charged leptons. It is therefore possible to construct very predictive models.
In a sufficiently predictive model, the quark and charged lepton masses and
CKM angles can be enough to determine $M_N$ completely. Such is the case in
the model of Ref. \cite{ab}, which is therefore a good laboratory for
studying leptogenesis. In that very predictive and realistic model
the Dirac neutrino mass matrix is given by
\begin{equation}
M_N = \left( \begin{array}{ccc}
\eta & 0 & 0 \\ 0 & 0 & \epsilon \\ 0 & -\epsilon & 1 \end{array}
\right) v_u,
\end{equation}

\noindent
where $\eta \cong 0.6 \times 10^{-5}$ and $\epsilon \cong 0.14$. The $SO(10)$
symmetry does not directly relate the matrices $F_{ij}$ and $F'_{ij}$ 
to anything known, and so there is considerable freedom in choosing their 
forms, and consequently also the form of the matrix $H_{ij}$. As pointed out 
in Ref. \cite{type3}, there are two simple cases which lead to realistic
patterns of masses and mixings for the light neutrinos $\nu_i$. 

In Case 1, all the elements of $F_{ij}$ and of $F'_{ij}$ are of order $f$, 
some dimensionless parameter. Then 
\begin{equation}
H_{ij} \sim \left( \begin{array}{ccc} 1 & 1 & 1 \\ 1 & 1 & 1 \\ 1 & 1 & 1
\end{array} \right),
\end{equation}

\noindent 
and therefore, by Eq. (7), the type III seesaw mass formula gives
\begin{equation} 
M_{\nu} \sim \left( \begin{array}{ccc} \eta & \epsilon & 1 \\
\epsilon & \epsilon & 1 \\ 1 & 1 & 1 \end{array} \right) \frac{v_u u}{\Omega}.
\end{equation}

\noindent
It is evident by inspection of the 12 and 23 elements of $M_{\nu}$ that the 
solar and atmospheric angles naturally come out of the right order. However, 
some very mild tuning (if one should even call it that)
of order $10^{-1}$ is needed to make the 13 element 
small enough to satisfy the experimental constraint on $U_{e3}$.  In Case 1 
all the right-handed neutrinos have mass of the same order, namely of
order $\Omega \sim M_{GUT}$. This is not good for thermal leptogenesis,
as it would require a reheating temperature of order $M_{GUT}$, leading to
the overproduction of gravitinos. 

In Case 2, both $F_{ij}$ and $F'_{ij}$ have the form
\begin{equation}
F,F' \sim \left( \begin{array}{ccc} \lambda^2 & \lambda & \lambda \\
\lambda & 1 & 1 \\ \lambda & 1 & 1 \end{array} \right),
\end{equation}

\noindent
where $\lambda \sim \eta/\epsilon$. This form can naturally arise from 
an abelian flavor symmetry under which the first family has a different 
charge from the other two. One would therefore assume 
that ${\cal M}_{ij}$ also has the form shown in Eq. (11). Such an abelian
flavor symmetry would also explain
why the element $\eta$ in Eq. (8) is so much smaller than the other
non-zero elements in $M_N$. From Eqs. (7) and (11),
\begin{equation}
H \sim \left( \begin{array}{ccc}
1 & \epsilon/\eta & \epsilon/\eta \\ \eta/\epsilon & 1 & 1 \\ 
\eta/\epsilon & 1 & 1 \end{array} \right),
\end{equation}

\noindent
and
\begin{equation}
M_{\nu} \sim \left( \begin{array}{ccc} \eta & \epsilon & \epsilon \\
\epsilon & \epsilon & 1 \\ \epsilon & 1 & 1 \end{array} 
\right) \frac{v_u u}{\Omega}.
\end{equation}

\noindent
Note that this naturally gives all three neutrino mixing angles
of the correct order: $|U_{\mu 3}| \sim 1$, $|U_{e2}| \sim 1$, and
$|U_{e3}| \sim \epsilon$. 

In the limit where the matrix ${\cal M}_{ij}$ can be
neglected, the superheavy neutrino masses are given simply by the matrix
$F_{ij} \Omega$. Therefore, in Case 2, two of the superheavy
(Dirac) neutrinos have mass of order $\Omega \sim M_{GUT}$ and one
superheavy (Dirac) neutrino has mass of order $\lambda^2 \Omega
\sim 10^7$ GeV, which is quite satisfactory from the point of view of
thermal leptogenesis. This is the case that we shall now consider in more
detail.

To analyze leptogenesis it is convenient to go to a basis in which the matrix 
$F_{ij}$ is diagonal. We shall indicate quantities in this basis with a
tilde. To reach this basis one does unitary transformations
$N^c_i = U_{ij} \tilde{N}^c_j$ and $S_i = V_{ij} \tilde{S}_j$. From the
assumed form of $F_{ij}$ in Eq. (11) we can write the matrix $U_{ij}$ as
\begin{equation}
U = \left( \begin{array}{ccc} u_{11} & \lambda u_{12} & \lambda u_{13} \\
\lambda u_{21} & u_{22} & u_{23} \\ \lambda u_{31} & u_{32} & u_{33} 
\end{array} \right),
\end{equation}

\noindent
where $u_{ij} \sim 1$, and by unitarity $u_{11} \cong 1$. The unitary
matrix $V_{ij}$ has a similar form; however, we shall not need
to parametrize it. The elements of the
diagonalized matrix $\tilde{F}_{ij} \Omega = (U_{ki} F_{k \ell} V_{\ell j})
\Omega$ can be written
\begin{equation}
\tilde{F} \Omega = \left( \begin{array}{ccc} \lambda^2 F_1 & 0 & 0 \\
0 & F_2 & 0 \\ 0 & 0 & F_3 \end{array} \right) \Omega \equiv 
\left( \begin{array}{ccc} M_1 & 0 & 0 \\ 0 & M_2 & 0 \\ 0 & 0 & M_3 
\end{array} \right),
\end{equation}

\noindent
where $F_i \sim 1$. In the new basis the matrices $\tilde{F}'_{ij} u$ and 
$\tilde{{\cal M}}_{ij}$ are not diagonal, but rather
retain the same basic form shown in Eq. (11), so that they can be
parametrized as follows:
\begin{equation}
\tilde{F}' u = \left( \begin{array}{ccc}
\lambda^2 f_{11} & \lambda f_{12} & \lambda f_{13} \\
\lambda f_{21} & f_{22} & f_{23} \\ \lambda f_{31} & f_{32} & f_{33} 
\end{array} \right) v_u, \;\;
\tilde{{\cal M}} = \left( \begin{array}{ccc}
\lambda^2 g_{11} & \lambda g_{12} & \lambda g_{13} \\
\lambda g_{21} & g_{22} & g_{23} \\ \lambda g_{31} & g_{32} & g_{33} 
\end{array} \right) M_S,
\end{equation}

\noindent
where $f_{ij}, g_{ij} \sim 1$. The magnitude of $M_S$ is assumed to be
much less than $\Omega$. As will be seen, this is necessary for sufficient
leptogenesis.

The Dirac neutrino mass matrix $M_N$ goes to
$\tilde{M}_N = M_N U$, so that from Eqs. (8) and (14) it is given by
\begin{equation}
\tilde{M}_N \cong \left( \begin{array}{ccc}
\eta u_{11} & \eta \lambda u_{12} & \eta \lambda u_{13} \\
\epsilon \lambda u_{31} & \epsilon u_{32} & \epsilon u_{33} \\
\lambda u_{31} & u_{32} & u_{33} \end{array} \right) v_u \equiv
\tilde{Y} v_u.
\end{equation}

The type III seesaw contribution to the light neutrino
mass matrix (which dominates, since $M_S \ll \Omega$) is determined from
Eq. (7) to be
\begin{equation}
M_{\nu} \cong - \left[ \begin{array}{ccc}
2 \eta \left( \frac{u_{11} f_{11}}{F_1} \right) & \frac{\eta}{\lambda} 
\left( \frac{u_{11} f_{21}}{F_1} \right) & \frac{\eta}{\lambda} 
\left( \frac{u_{11} f_{31}}{F_1} \right) \\ \frac{\eta}{\lambda} 
\left( \frac{u_{11} f_{21}}{F_1} \right) & 2 \epsilon \sum_j
\left( \frac{u_{3j} f_{2j}}{F_j} \right) & 
\sum_j \left( \frac{u_{3j} f_{2j}}{F_j} \right) \\
\frac{\eta}{\lambda} \left( \frac{u_{11} f_{31}}{F_1} \right)
& \sum_j \left( \frac{u_{3j} f_{2j}}{F_j} \right) & 
2 \sum_j \left( \frac{u_{3j} f_{3j}}{F_j} \right) \end{array}
\right] \left( \frac{v_u^2}{\Omega} \right).
\end{equation}

\noindent
Since $\lambda$ is presumed to be of order $\eta/\epsilon$, one sees that
$M_{\nu}$ does indeed have the form given in Eq. (13). 

The six superheavy two-component neutrinos have the mass matrix
\begin{equation} 
(\tilde{N}^c_i, \tilde{S}_i) \left( \begin{array}{cc}
0 & M_i \delta_{ij} \\ M_i \delta_{ij} & \tilde{{\cal M}}_{ij} \end{array}
\right) \left( \begin{array}{c} \tilde{N}^c_j \\ \tilde{S}_j \end{array} 
\right), 
\end{equation}

\noindent
where $\tilde{{\cal M}}_{ij}$ is given in Eq. (16). Leptogenesis is 
almost exclusively produced by the decays of the lightest pair of these 
superheavy neutrinos, which have effectively the two-by-two mass matrix
\begin{equation} 
(\tilde{N}^c_1, \tilde{S}_1) \left( \begin{array}{cc}
0 & M_1  \\ M_1  & \tilde{{\cal M}}_{11} \end{array}
\right) \left( \begin{array}{c} \tilde{N}^c_1 \\ \tilde{S}_1 \end{array} 
\right) =
(\tilde{N}^c_1, \tilde{S}_1) \; \lambda^2 \left( \begin{array}{cc}
0 & F_1 \Omega  \\ F_1 \Omega  & g_{11} M_S \end{array}
\right) \left( \begin{array}{c} \tilde{N}^c_1 \\ \tilde{S}_1 \end{array} 
\right).
\end{equation}

\noindent
If, as we assume, $M_S \ll \Omega$, these form an almost degenerate 
pseudo-Dirac pair, or equivalently two Majorana neutrinos with nearly equal
and opposite masses. These Majorana neutrinos are 
$N_{\pm} \cong (\tilde{N}^c_1 \pm \tilde{S}_1)/\sqrt{2}$, with masses
$M_{\pm} \cong \pm M_1 + \frac{1}{2} \tilde{{\cal M}}_{11} =
\lambda^2 (\pm F_1 \Omega + \frac{1}{2} g_{11} M_S)$.
These can decay into light neutrino plus Higgs via the term
$Y_{i \pm}(N_{\pm} \nu_i)H$, where
\begin{equation}
Y_{i \pm} \cong (\tilde{Y}_{i1} \pm \tilde{F}'_{i1})/\sqrt{2}
\mp \frac{\tilde{{\cal M}}_{11}}{4 M_1}
(\tilde{Y}_{i1} \mp \tilde{F}'_{i1})/\sqrt{2}.
\end{equation}

\noindent
Here $\tilde{Y}$ is the Dirac Yukawa coupling matrix given in Eq. (17).
It is straightforward to show that the lepton asymmetry per decay
produced by the decays of $N_{\pm}$ is given (in the now usual notation)
by \cite{leptogen2,leptogen3,leptogen4}
\begin{equation}
\epsilon_1 = \frac{1}{4 \pi} \frac{Im [\sum_j(Y_{j+} Y^*_{j-})]^2}
{\sum_j [|Y_{j+}|^2 + |Y_{j-}|^2]} I(M^2_-/M^2_+),
\end{equation}

\noindent
where $I(M^2_-/M^2_+)$ comes from the absorptive part of the decay amplitude
of $N_{\pm}$. This function is given by $I(x) = \sqrt{x}[1 - 
(1+x) \ln (1+ \frac{1}{x}) + \frac{1}{1-x}]$. Thus $I(M^2_-/M^2_+) \cong
M_1/2\tilde{{\cal M}}_{11} = (F_1/2 g_{11})(\Omega/M_S)$. We will call this
the resonant enhancement factor. It can be large if $M_S \ll \Omega$.
The expression for $I(M^2_-/M^2_+)$ given above is only valid when the 
mass splitting $|M_+| - |M_-| = \tilde{{\cal M}}_{11}$ is larger than the 
widths of the $N_{\pm}$, which are given by 
$\Gamma_{\pm} \cong \frac{1}{8 \pi} M_1
\sum_k|Y_{k \pm}|^2$. From Eqs. (16), (17), and (21), one sees that 
$\Gamma_{\pm} \sim \lambda^2 M_1/8 \pi$. Thus the condition for maximum
lepton asymmetry resulting from resonant enhancement, that the splitting
of $N_{\pm}$ be comparable to their widths \cite{rescond}, only constrains 
the enhancement factor to be less than about
$8 \pi/\lambda^2 \sim 10^{10}$. This is not a problem as we shall only
need enhancements of order $10^5$ or so.

Making use of Eqs. (21) and (22) one obtains
\begin{equation}
\epsilon_1 = \frac{1}{4 \pi} \frac{\sum_j(|\tilde{Y}_{j1}|^2 - 
|\tilde{F}'_{j1}|^2) Im(\sum_k \tilde{Y}^*_{k1} \tilde{F}'_{k1})}
{\sum_j (|\tilde{Y}_{j1}|^2 + |\tilde{F}'_{j1}|^2)} I(M^2_-/M^2_+),
\end{equation}
 
\noindent
This can be evaluated in terms of the parameters of the model using Eqs. (16) 
and (17), giving
\begin{equation}
\epsilon_1 \cong \frac{\lambda^2}{4 \pi} \left[ \frac{(|u_{31}|^2 -
|f_{31}|^2) Im(u_{31}^* f_{31})}{|u_{31}|^2 + |f_{31}|^2 + |f_{21}|^2}
\right] I.
\end{equation}

\noindent
The lepton asymmetry of the universe is given by 
\cite{leptogen2,leptogen3,leptogen4}
\begin{equation}
Y_L = \epsilon_1 d/g_*,
\end{equation}

\noindent
where $d$ is the washout parameter and $g_*$ is the effective number of
polarization states at the time of the $N_{\pm}$ decays. In a supersymmetric
model $g_* = 228.75$. The washout parameter is given approximately by
\cite{washout}
\begin{equation}
d = \frac{0.3}{k (\ln k)^{3/5}}, \;\; k \equiv \frac{\tilde{m}_1}
{10^{-3} {\rm eV}},
\end{equation}

\noindent
where
\begin{equation}
\tilde{m}_1 \equiv \frac{8 \pi v_u^2 \Gamma_{N_{\pm}}}{M^2_{N_{\pm}}}
\cong \lambda^2 \frac{v_u^2}{M_1} (|u_{31}|^2 + |f_{31}|^2 + |f_{21}|^2).
\end{equation}

Before further analyzing leptogenesis in this model, it is useful to find the
constraints imposed on its parameters by the condition that realistic
masses and mixings arise for the light neutrinos. The two conditions that
are important are that $m_3$ (the largest of the light neutrino
masses) and $\theta_{12}$ (the solar neutrino mixing angle) come out
right. One expects that $m_3 \simeq (M_{\nu})_{33} = 2 \sum_j (u_{3j}
f_{3j}/F_j)(v^2_u/\Omega)$, cf. Eq. (18). It will turn out to be convenient to 
express this in terms of the following parameter:
\begin{equation}
a \equiv \sum_j \left[\frac{u_{3j} f_{3j}}{|f_{21}|^2} \frac{F_1}{F_j} \right],
\end{equation}

\noindent
which is naturally of order one. Then one requires that
\begin{equation}
m_3 \simeq \frac{2 |f_{21}|^2}{F_1} a \frac{v_u^2}{\Omega} \simeq 0.06 \; 
{\rm eV}.
\end{equation}

\noindent
Since $2 v_u^2/m_3 \simeq 10^{15}$ GeV, 
\begin{equation}
\Omega \simeq \frac{|f_{21}|^2 a}{F_1} [10^{15} {\rm GeV}].
\end{equation}

\noindent
Using the fact that $M_1 = \lambda^2 F_1 \Omega$ (cf. Eq. (15)), this can also
be written in the following form, which shall be useful shortly:
\begin{equation}
\frac{\lambda^2}{M_1} \simeq \frac{1}{|f_{21}|^2 a} (10^{15} {\rm GeV})^{-1}.
\end{equation}

The condition that the solar angle satisfy $\tan^2 \theta_{12} \simeq 0.4$
implies that $(M_{\nu})_{12} \simeq (M_{\nu})_{22}$. From
the fact that $m_2 \simeq m_3/6$ (if the light neutrino masses have an
ordinary hierarchy), it follows that $(M_{\nu})_{12} \simeq 0.1 \; m_3$ and
hence, from Eq. (18), that $(\eta/\lambda)(|f_{21}|/F_1)(v_u^2/\Omega)
\simeq 0.006 \; {\rm eV}$. Again using the formula for $M_1$ and the value
of $\eta$, one 
arrives at the result that
\begin{equation}
\frac{\lambda}{M_1} \simeq \frac{1}{|f_{21}|} (3 \times 10^{10} 
	{\rm GeV})^{-1}.
\end{equation}

\noindent
Dividing Eq. (31) by the square of Eq. (32) gives
\begin{equation}
M_1 \simeq \frac{1}{a} (0.9 \times 10^6 {\rm GeV}).
\end{equation}

\noindent
This is certainly small enough for satisfactory thermal leptogenesis.
Dividing Eq. (31) by Eq. (32) yields
\begin{equation}
\lambda \simeq \frac{1}{|f_{21}| a} (3 \times 10^{-5}),
\end{equation}

\noindent
which indeed is of order $\eta/\epsilon = 4 \times 10^{-5}$. For ease of 
writing, let us define $x \equiv f_{31}/|f_{21}|$ and $y \equiv u_{31}/
|f_{21}|$. Then combining Eqs. (27) and (31) yields
\begin{equation}
\tilde{m}_1 \simeq \frac{1}{a} (3 \times 10^{-2} {\rm eV}) (1 + |x|^2 + |y|^2).
\end{equation}

\noindent
Assembling Eqs. (24), (25), (26), (34) and (35) (and using $(\ln k)^{3/5}
\simeq 3$, which is a reasonable approximation for the values of
$\tilde{m}_1$ that will be of interest) one has finally the result
\begin{equation}
Y_L \simeq \frac{10^{-15}}{a} \left[ 
\frac{(|x|^2 - |y|^2)Im(x^*y)}{(1 + |x|^2 + |y|^2)^2} \right] I.
\end{equation}

\noindent
The maximum value of the function of $x$ and $y$ in the brackets is $1/4$.
Thus sufficient leptogenesis requires a resonant enhancement factor
of about $I \sim 4 \times 10^5$, well below the upper bound of $10^{10}$ 
imposed to ensure that the mass splitting of the quasi-Dirac pair is greater 
than their widths. If $a$ happens to be small, a smaller 
enhancement is needed, but for $a$ to be much less than one would be a 
fine-tuning. 

\section{Comparison with leptogenesis in type I seesaw}

To illustrate the advantages of the type III seesaw for both leptogenesis
and obtaining realistic light neutrino masses, we shall briefly examine
these issues in the realistic $SO(10)$ model that we have been using as
a laboratory. Let us therefore now suppose that in that model only the
conventional type I seesaw mechanism operates. It will be useful to
parametrize the inverse of the matrix $M_R$ as follows:
\begin{equation}
M^{-1}_R = \left[ \begin{array}{ccc} a \left( \frac{\epsilon}{\eta} \right)^2 
& b \left( \frac{1}{\eta} \right) & c \left( \frac{\epsilon}{\eta} \right)
\\ b \left( \frac{1}{\eta} \right) & d \left( \frac{1}{\epsilon} \right)^2
& e \left( \frac{1}{\epsilon} \right) \\ c \left( \frac{\epsilon}{\eta} \right)
& e \left( \frac{1}{\epsilon} \right) & 1 \end{array} \right] m^{-1}_R,
\end{equation}

\noindent
where in order to get a realistic $M_{\nu}$, we take $a$, $b$, $c$, $d$, and
$e$ to be of 
order one. Note that this matrix has a geometric hierarchy of the type that
typically arises from an abelian flavor symmetry. From the form of $M_N$
given in Eq. (8) and the type I seesaw formula $M_{\nu} = - M_N M^{-1}_R
M_N^T$, one has
\begin{equation}
M_{\nu} = - \left[ \begin{array}{ccc} a \epsilon^2 &  c \epsilon^2 &
(c-b) \epsilon \\ c \epsilon^2 & \epsilon^2 & (1-e) \epsilon \\
(c-b) \epsilon & (1-e) \epsilon & (1 - 2e + d) \end{array} \right] 
\frac{v^2_u}{m_R}.
\end{equation}

\noindent
The first thing to notice is that the ratio
of the two larger light neutrino masses, $m_2/m_3$, is of order
$\epsilon^2$. However, empirically this ratio is roughly equal to $\epsilon 
\cong 0.14$. Thus there must be a mild fine-tuning to make $(1-2e+d) 
= O(\epsilon)$. 

A much more precise tuning is required to have a large resonant
enhancement of leptogenesis. Suppose that a resonant enhancement of 
$I \gg 1$ is required. Then the two lightest ``right-handed" neutrinos,
$N^c_1$ and $N^c_2$, must form a pseudo-Dirac pair with a mass splitting 
$\Delta M /M = 1/(2I)$. 
Examining the 12 block of the matrix $M_R^{-1}$ in Eq. (37)
one sees that this requires that $\left[ d \left(\frac{1}{\epsilon} \right)^2
+ a \left( \frac{\epsilon}{\eta} \right)^2 \right] = \frac{1}{2I} b \left( 
\frac{1}{\eta} \right)$. That is,
$a/b = [\eta/(2 \epsilon^2)] I^{-1} = 1.5 \times 10^{-4} I^{-1}$.

A resonant enhancement of order $10^4$, then, would require a 
tuning of the parameter $a$ to be of order $10^{-8}$. This tuning of the 
11 component
of $M_R^{-1}$ is obviously equivalent to the tuning of the determinant of 
the 23 block of $M_R$. In Ref. \cite{abtype1} it is argued that in a particular
version of this model (cf. \cite{abdetailed}) this subdeterminant of $M_R$
may have a natural reason to be small compared to one, though not
as small as $10^{-8}$. 

We have considered a specific form of $M_N$ here that comes from a particular
model of quark and lepton masses. However, this model is a fairly typical
one, and it seems likely that a similar extreme fine-tuning of the form of 
$M_R$ would be required in a wide class of realistic models having the 
type I seesaw in order to produce large resonant enhancement of leptogenesis.
This is the ``clash" mentioned in the Introduction between the requirements
of realistic $M_{\nu}$ and leptogenesis in type I seesaw models.

In the type III seesaw we found that a large resonant enhancement of 
leptogenesis required
that the mass $M_S$ that sets the scale for the matrix ${\cal M}$
must be small compared to $M_{GUT}$. It would not be hard to make this natural
by an abelian symmetry under which the singlets $S_i$ were charged.
In any case, the setting of a mass scale such as $M_S$ to be small seems 
less unnatural than having a highly tuned relationship among the elements 
of the matrix $M_R$.

\vspace*{0.3in}

The research of SMB was supported in part by Department of Energy Grant
Number DE FG02 91 ER 40626 A007.  One of us (CHA) thanks the Fermilab
Theoretical Physics Department for its kind hospitality where his work
was carried out.  Fermilab is operated by Universities Research Association
Inc. under contract with the Department of Energy.

\newpage

\noindent

\begin{picture}(360,216)
\thicklines
\put(36,108){\vector(1,0){36}}
\put(72,108){\line(1,0){36}}
\put(108,72){\line(0,1){36}}
\put(103,69){$\times$}
\put(108,108){\line(1,0){36}}
\put(180,108){\vector(-1,0){36}}
\put(177,105){$\times$}
\put(180,108){\vector(1,0){36}}
\put(216,108){\line(1,0){36}}
\put(252,72){\line(0,1){36}}
\put(247,69){$\times$}
\put(252,108){\line(1,0){36}}
\put(324,108){\vector(-1,0){36}}
\put(18,117){$N^c_i=1(16_i)$}
\put(100,117){$F^a_{ik}$}
\put(90,48){$\langle 1(\overline{16}^a_H) \rangle \equiv \Omega_a$}
\put(129,129){$S_k =$}
\put(129,117){$1(1_k)$}
\put(168,90){${\cal M}_{k \ell}$}
\put(199,129){$S_{\ell}=$}
\put(199,117){$1(1_{\ell})$}
\put(245,117){$F^b_{j \ell}$}
\put(234,48){$\langle 1(\overline{16}^b_H)\rangle \equiv \Omega_b$}
\put(288,117){$N^c_j = 1(16_j)$}
\end{picture}

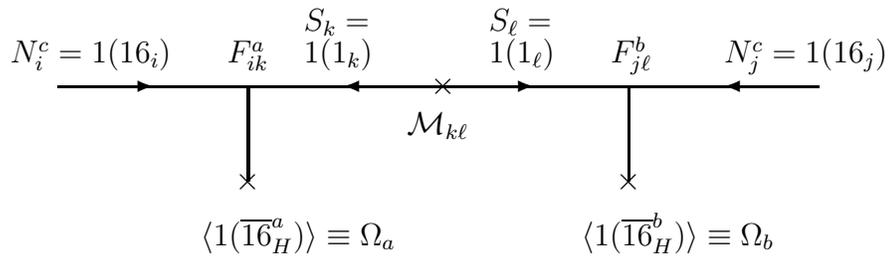
Fig. 1. \begin{minipage}[t]{5in}{Diagram that produces the effective 
operator ${\bf 16}_i {\bf 16}_j \overline{{\bf 16}}^a_H 
\overline{{\bf 16}}^b_H/M_G$, which generates $M_R$.}\end{minipage}

\end{document}